\documentclass[12pt]{iopart}
\pdfoutput=1
\expandafter\let\csname equation*\endcsname\relax
\expandafter\let\csname endequation*\endcsname\relax
\usepackage[numbers,sort&compress]{natbib}
\usepackage{amsmath,amssymb,eucal,graphicx,bm}
\usepackage{subfigure}
\usepackage{float}

\def\ltwid{\mathrel{\raise.3ex\hbox{$<$\kern-.75em\lower1ex\hbox{$\sim$}}}}

\usepackage{comment}
\usepackage{appendix}
\usepackage[breakwords]{truncate}
\usepackage{lastpage}
\usepackage{hyperref}
\usepackage{amsfonts}
\usepackage{physics}
\usepackage[section]{placeins}
\usepackage{color,url}

\begin{document}

\title{First-Passage-Driven Boundary Recession}
\author{B. De Bruyne}
\address{LPTMS, CNRS, Univ.\ Paris-Sud, Universit\'e Paris-Saclay, 91405 Orsay, France}
\author{J. Randon-Furling}
\address{SAMM, Universit\'e Paris 1 -- FP2M (FR2036) CNRS, 75013 Paris, France}
\address{MSDA, Mohammed VI Polytechnic University, Ben Guerir 43150, Morocco}
\author{S. Redner}
\address{Santa Fe Institute, 1399 Hyde Park Rd., Santa Fe, New Mexico 87501, USA}

\begin{abstract}
  We investigate a moving boundary problem for a Brownian particle on the
  semi-infinite line in which the boundary moves by a distance proportional
  to the time between successive collisions of the particle and the boundary.
  Phenomenologically rich dynamics arises.  In particular, the probability
  for the particle to first reach the moving boundary for the $n^\text{th}$
  time asymptotically scales as $t^{-(1+2^{-n})}$.  Because the tail of this
  distribution becomes progressively fatter, the typical time between
  successive first passages systematically gets longer.  We also find that the
  number of collisions between the particle and the boundary scales as
  $\ln\ln t$, while the time dependence of the boundary position varies as
  $t/\ln t$.
\end{abstract}

\section{Introduction and Model}

Moving boundary problems arise in materials that are near a first-order phase
transition (see, e.g.,
\cite{crank1984free,rubinstein2000stefan,RevModPhys.52.1} for general
introductions).  Perhaps the most familiar examples are the melting of ice
that is immersed in water, or the freezing of water on the surface of a lake
when the ambient air suddenly cools to a temperature $T< 0^\circ$C at some
initial time $t=0$.  In the latter case, a layer of ice starts growing on top
of the water.  The lower ice-water interface remains at $0^\circ$C, while
heat is conducted to the upper ice-air interface and thence into the air.  If
the ice-air interface is defined to be at spatial position $z=0$ and the
ice-water interface is at $-L(t)$, then to a first approximation, the
temperature in the ice at vertical position $z$ is $T(z)=Tz/L$.  As a result
of this temperature gradient and the resulting heat conduction, molecules of
water at the interface freeze and join the ice layer.  Through this
mechanism, the interface gradually grows downward at a rate that is
proportional to the temperature gradient in the ice; this gives
$L(t)\sim\sqrt{t}$.

Microscopically, heat conduction corresponds to the diffusion of phonons,
with interface motion occurring when the phonons first reach the interface.
It is in this sense that we can think of the motion of the boundary being
controlled by a first-passage process.  Namely, whenever a diffusing particle
reaches the interface, the interface then moves by a specified amount.
In~\cite{de2021optimization}, we studied an idealization of this problem in a
strictly one-dimensional geometry for the two cases in which the interface
recedes by a fixed distance or multiplicatively, whenever a diffusing
particle reaches the interface.  After each collision between the particle
and the interface, the particle is returned to its starting position and the
process begins anew.  This process defines a simple first-passage resetting
problem.  This is a natural complement to Poisson resetting, where a random
walk or a diffusing particle is returned to its starting position at some
fixed
rate\cite{evans2011diffusion,evans2011diffusionjpa,evans2020stochastic,PhysRevLett.112.240601,christou2015diffusion,PhysRevE.92.060101,PhysRevE.92.052126,reuveni2016optimal,pal2017first,belan2018restart,bodrova2019nonrenewal}.
The consequences of Poisson resetting have been extensively investigated, but
first-passage resetting is much less explored thus far.

In the former case where the boundary moves by a fixed distance after each
first-passage event~\cite{de2021optimization}, we found that the number of
collisions between the particle and interface, as well as the position of the
interface grew as $t^{1/4}$.  This should be compared to the $t^{1/2}$ growth
of the number of collisions when there is no resetting of the particle
position.  In the case of multiplicative motion, the interface, which is
initially located at $L$, moves to $\alpha L$, $\alpha^2 L$, $\alpha^3 L$,
after each successive collision. Because this interface motion is rapid, the
number of collisions now grows only logarithmically in time.  In spite of the
small number of resetting events, the interface position still grows as
$t^{1/2}$ because later collisions between the particle and interface leads
to a large displacement of the interface.

\begin{figure}[ht]
  \begin{center}
    \includegraphics[width=0.45\textwidth]{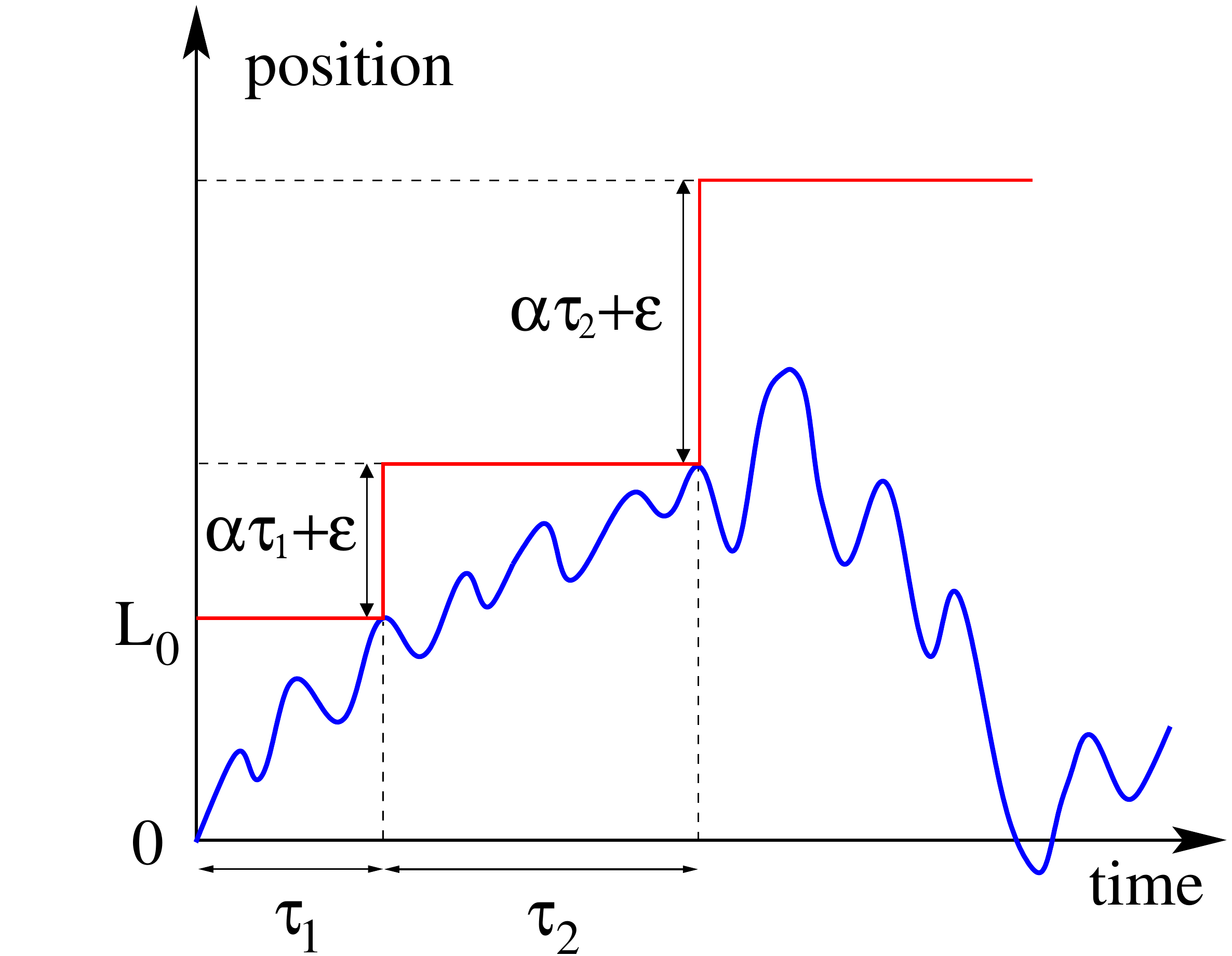}
    \caption{Illustration of the moving boundary (red line) in a
      semi-infinite geometry due to first-passage resetting.  Each time the
      particle reaches the boundary, the boundary recedes by a distance
      $\alpha \tau_i+\epsilon$. The successive first-passage times are
      denoted $\tau_1,\tau_2,\ldots$.}
    \label{fig:model}
  \end{center}
\end{figure}

In this work, we study another natural scenario for first-passage resetting
where the boundary recedes by an amount that is proportional to the time
difference between successive encounters of the particle and the interface
(figure~\ref{fig:model}).  Thus we investigate the dynamics of a
one-dimensional Brownian particle with diffusion coefficient $D$ that starts
at the origin in a semi-infinite geometry, with a boundary that is initially
a distance $L_0$ from the particle.  This boundary remains fixed except for
these instances when the particle reaches it.  When such an encounter
happens, the boundary moves away from the particle by a distance that is
proportional to the first-passage time between the previous and the current
encounter.  While there is no obvious physical motivation for this model, the
resulting phenomenology is quite rich and perhaps has some fundamental
implications.

As illustrated in figure~\ref{fig:model}, it is necessary to augment the
recession distance of the interface by a small additive amount to regularize
a singular behavior that arises when there is no cutoff.  If the interface
recedes by an amount that is strictly linear in the first-passage time from
the previous collision, a pathology arises in which the particle can hit the
interface infinitely often in a finite time.  That is, if the first-passage
time is short, the time to the next collision can be even shorter, ultimately
leading to a singularity.  For a random walk on a lattice, there is no such
pathology because the minimum first-passage time cannot be less than the time
for a single step.  To obviate the pathology in the case of continuum
diffusion, we define the recession distance of the interface to be equal to
the first-passage time from the previous collision plus a small cutoff
$\epsilon$.  Our results are independent of this cutoff, but this cutoff is
necessary to obtain non-singular results.

Our primary results are the following: (a) The probability for the particle
to first reach the interface for the $n^{\text th}$ time asymptotically
decays as $t^{-\beta_n}$, with $\beta_n= 1+2^{-n}$.  Thus the tail of the
$n^\text{th}$-passage probability becomes progressively fatter after each
collision. (b) The number of collisions between the particle and the
interface grows with time as $\ln\ln t$.  This slow double logarithmic
increase in the number of collisions is reminiscent of the intriguing
Khintchine iterated logarithm
law~\cite{khintchine1924,feller2008introduction,morters2010brownian} for the
extreme position of Brownian motion.  Finally, the position of the interface
recedes according to $t/\ln t$.  Once again, even though collisions between
the particle and the boundary are rare, the widely separated collisions in
time lead to a large displacement of the boundary, so that its overall motion
is nearly ballistic.

\section{Successive-Passage Distributions}

\subsection{The second-passage distribution}\label{sec:lap2}


We define $F_n(t\,|\,L_0)$ as the probability that a diffusing particle,
which starts at $x=0$, \emph{first} reaches the boundary at $x=L_0$ for the
$n^\text{th}$ time at time $t$.  When $n=1$, this quantity is just the
first-passage probability distribution for a Brownian motion to reach
$x=L_0$~\cite{redner2001guide,bray2013persistence}:
\begin{align}
\label{eq:f1}
F_1(t\,|\,L_0) \equiv F(t\,|\,L_0)= \frac{L_0}{\sqrt{4\pi Dt^3}}\;e^{-L_0^2/4Dt}\,.
\end{align}
Because of the convolution structure of the problem we are treating, it is
convenient to work in the Laplace domain.  Thus we also introduce the Laplace
transform of $F_1(t\,|\,L_0)$:
\begin{align}
  \widetilde F_1(s\,|\,L_0) = \int_0^\infty dt\, e^{-st} F_1(t\,|\,L_0)
  = \,e^{-L_0\sqrt{s/D}}\equiv e^{-L_0\, g_1(s)}\,,\label{eq:lapf1}
\end{align}
with $g_1(s)=\sqrt{s/D}$.

For the particle to first reach the boundary for the second time at time $t$,
it must first reach the boundary at an intermediate time $\tau_1$ and then
reach it one more time during the remaining time $\tau_2$, with
$t=\tau_1+\tau_2$. Since the boundary will have moved by a distance
$\tau_1+\epsilon$ after the first encounter (see figure~\ref{fig:model}), the
second-passage distribution is given by
\begin{align}
  F_2(t\,|\,L_0) &= \int_0^{\infty} d\tau_1\int_0^{\infty} d\tau_2\, F_1(\tau_1\,|\,L_0) \times F(\tau_2\,|\,\alpha\, \tau_1+\epsilon)\,\delta(t-\tau_1-\tau_2)\,,\label{eq:f2}
\end{align}
Taking the Laplace transform gives
\begin{subequations}
\begin{align}
  \widetilde F_2(s\,|\,L_0) &= \int_0^{\infty} d\tau_1\int_0^{\infty} d\tau_2\, F_1(\tau_1\,|\,L_0)  F(\tau_2\,|\,\alpha\, \tau_1+\epsilon)\,e^{-s\tau_1-s\tau_2}\,,\nonumber\\[1mm]
   &=\int_0^{\infty} d\tau_1\, F_1(\tau_1\,|\,L_0) \, e^{-s\tau_1}
\int_0^{\infty} d\tau_2\, F(\tau_2\,|\,\alpha\, \tau_1+\epsilon)\,e^{-s\tau_2}\,,\label{eq:f2s}
\end{align}
where the second integral is simply the Laplace transform of the
first-passage distribution \eqref{eq:lapf1} with
$L_0=\alpha \tau_1+\epsilon$.  Using this fact, we obtain
\begin{align}
  \widetilde F_2(s\,|\,L_0)&=\int_0^{\infty} d\tau_1\, F_1(\tau_1\,|\,L_0)\,  e^{-s\tau_1}\;
   e^{-(\alpha\,\tau_1+\epsilon)\sqrt{s/D}}\,,\nonumber\\
  &=e^{-\epsilon \sqrt{s/D}}\int_0^{\infty} d\tau_1\, F_1(\tau_1\,|\,L_0) e^{-\tau_1\left(s+\alpha\sqrt{s/D}\right)}\label{eq:f2s2}
\end{align}
The integral is just the Laplace transform of the first-passage distribution
\eqref{eq:lapf1}, but now evaluated at $s+\alpha\sqrt{s/D}$.  Thus the
second-passage probability is
\begin{align}
  \label{eq:f2s3}
  \widetilde F_2(s\,|\,L_0)= \exp\bigg[-\epsilon \sqrt{s/D}
                     -L_0\frac{\sqrt{s+\alpha\sqrt{{s}/{D}}}}{\sqrt{D}}\bigg]
\equiv e^{ -\epsilon \sqrt{s/D}-L_0\, g_2(s)}      \,,
\end{align}
\end{subequations}
where
\begin{align*}
  g_2(s) = \frac{\sqrt{s+\alpha\sqrt{s/D}}}{\sqrt{D}} = \frac{\sqrt{s+\alpha
  g_1(s)}}{\sqrt{D}}\,.
\end{align*}
This expression for $ \widetilde F_2(s\,|\,L_0)$ cannot be Laplace inverted
analytically; however, in the relevant $s\to 0$ limit, its leading behavior
simplifies to
\begin{align*}
  \widetilde F_2(s\,|\,L_0)\to
  \exp\left[-L_0\,\sqrt{\frac{\alpha}{D}}\left(\frac{s}{D}\right)^{1/4}\right]\,,
\end{align*}
which can be Laplace inverted.  The result of this Laplace inversion is
\begin{align}
&\frac{L_0}{96 t^{5/4}} \left\{\frac{24}{\Gamma \left(\frac{3}{4}\right)} \; _0F_2\left(;\frac{1}{2},\frac{3}{4};\frac{L_0^4}{256
  {t}}\right)
  - \sqrt{\frac{288}{\pi }}\frac{L_0}{t^{1/4}} \,
   _0F_2\left(;\frac{3}{4},\frac{5}{4};\frac{L_0^4}{256
  {t}}\right) \right. \nonumber\\
 &\qquad\qquad\qquad \left.+  \frac{3 L_0^3}{ t^{1/2}\Gamma \left(\frac{5}{4}\right)} \; _0F_2\left(;\frac{5}{4},\frac{3}{2};\frac{L_0^4}{256
   {t}}\right)-\frac{\sqrt{8}\,L_0^3}{\pi t^{3/4}} \,
   _1F_3\left(1;\frac{5}{4},\frac{3}{2},\frac{7}{4};\frac{L_0^4}{256
   {t}}\right)\right\}\,,
\end{align}
where $_pF_q$ is the hypergeometric function.  The salient feature of this
formidable-looking expression is that the second-passage probability
asymptotically decays as $t^{-5/4}$.

\subsection{The $n^\text{th}$-passage distribution}\label{sec:lapn}

By repeatedly applying the reasoning that led to \eqref{eq:f2s3}, the Laplace
transform of the $n^\text{th}$-passage probability is
\begin{align}
\label{eq:fns}
\widetilde F_n(s\,|\,L_0)= \exp\bigg[-\epsilon \sum_{m=1}^{n-1}g_m(s)-L_0\, g_n(s)\bigg]\,,
\end{align}
where, for $n\geq 2$, $g_n(s)$ is defined by the recursion
\begin{align}
\label{eq:gns}
  g_n(s) = \frac{\sqrt{s+\alpha\, g_{n-1}(s)}}{\sqrt{D}}\,.
\end{align}

\begin{figure}[ht]
\begin{center}
\includegraphics[width=0.5\textwidth]{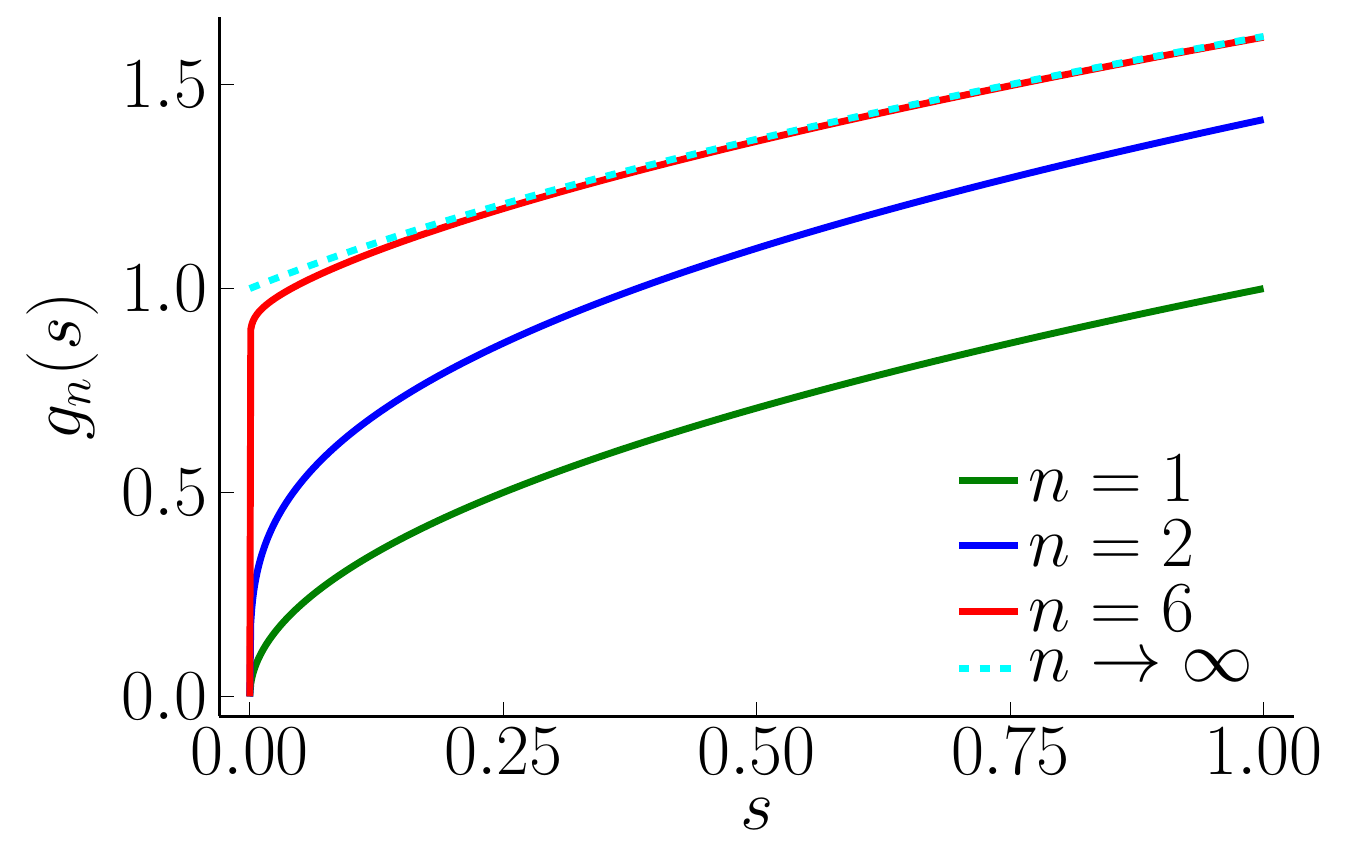}
\caption{The function $g_n(s)$ as a function of $s$ for different values of
  $n$ with $\alpha=D=1$. }
\label{fig:gns}
\end{center}
\end{figure}

The analytical structure of $g_n(s)$ provides the key to understanding the
problem.  In the limit $n\rightarrow \infty$ and $s$ fixed,
$g_n(s) \rightarrow g_\infty(s)$, where $g_\infty(s)$ satisfies
$g_\infty(s) = \sqrt{(s+\alpha\, g_\infty(s))/D}$.  The solution for
$g_\infty(s)$ is
\begin{align}
  g_\infty(s) =\frac{1}{2}\left(\frac{\alpha}{D} + \sqrt{\frac{\alpha^2}{D^2}+\frac{4s}{D}}\right)\,.\label{eq:gsso}
\end{align}
The function $g_n(s)$ is shown in figure \ref{fig:gns} for several values of
$n$, as well as for $n=\infty$.  Notice that $g_n(s)\to 0$ as $s\to 0$ for
any finite $n$, while $g_\infty(s\!=\!0)=\alpha/D$.  Also, when $\alpha=0$,
\eqref{eq:fns} reduces to the $n^\text{th}$-resetting probability
distribution that we obtained previously~\cite{de2021optimization}.

For $s\to 0$ and fixed $n$, the recurrence \eqref{eq:gns} reduces to
\begin{align}
  \label{eq:gns0}
  g_n(s) \sim \left(\frac{\alpha}{D}\right)^{1-2^{1-n}}\;
  \left(\frac{s}{D}\right)^{2^{-n}}\,.
\end{align}
Thus $g_n(s)$ becomes more singular as $n$ increases in such a way that
$g_\infty(0)=\alpha/D$.  We now substitute the above limiting behavior in
\eqref{eq:fns}, to give the small-$s$ expansion of the $n^\text{th}$ passage
distribution:
\begin{subequations}
\begin{align}
  \widetilde F_n(s\,|\,L_0)\sim
  1-L_0\left(\frac{\alpha}{D}\right)^{1-2^{1-n}}
  \left(\frac{s}{D}\right)^{2^{-n}}\,,\qquad s\rightarrow 0\,.\label{eq:fns0}
\end{align}
By a Tauberian theorem (for a simple derivation see Appendix A.2
of~\cite{evans2006canonical}), this gives the power-law decay in the time
domain
\begin{align}
\label{eq:fntl}
F_n(t\,|\,L_0) &\sim \frac{L_0\,\alpha^{1-2^{1-n}}}{-\Gamma(-2^{-n})D^{1-2^{-n}}}\,\frac{1}{t^{1+2^{-n}}}\,,\qquad t\rightarrow\infty\,.
\end{align}
\end{subequations}
Thus the long-time tail of the $n^\text{th}$-passage probability decays
progressively more slowly as $n$ increases, with the exponent value
approaching $-1$ for $n\to\infty$.


\section{Number of Encounters with the Boundary}

Because the tail of the $n^\text{th}$-passage probability becomes fatter as
$n$ increases, the times between encounters also become progressively
longer.  Thus we might expect that the total number of encounters will
increase only quite slowly with time.  We now show that this naive
expectation is what actually occurs.  Let $N(t)$ denote the number of times
that the particle reaches the boundary at time $t$.  The probability that
there are exactly $n$ encounters at time $t$ is formally given by
\begin{align}
  \label{eq:Pn}
P(N(t)\!=\!n\,|\,L_0) &= P(N(t)\!\geq\! n\,|\,L_0) - P(N(t)\!\geq\! n+1\,|\,L_0)\,,\nonumber\\
  &= \int_0^t d\tau F_n(\tau\,|\,L_0) -  \int_0^t d\tau F_{n+1}(\tau\,|\,L_0)\,.
\end{align}
The average number of encounters $\langle N(t) \rangle $ is given by
\begin{subequations}
\begin{align}
\label{eq:avgn}
\langle N(t) \rangle = \sum_{n=0}^\infty n\,P(N(t)\!=\!n\,|\,L_0)\,.
\end{align}
We substitute in $P(N(t)\!=\!n\,|\,L_0)$ from Eq.~\eqref{eq:Pn}, exploit the
telescopic nature of the sum, and also note that the $n=0$ gives no
contribution, to obtain
\begin{align}
  \label{eq:avgn2}
  \langle N(t) \rangle = \sum_{n=1}^\infty  P(N(t)\!\geq\! n\,|\,L_0)
  = \sum_{n=1}^\infty \int_0^t d\tau F_n(\tau\,|\,L_0)\,.
\end{align}
\end{subequations}
In the Laplace domain, the above relation becomes
\begin{subequations}
\begin{align}
  \label{eq:avgn3}
  \langle \widetilde N(s) \rangle = \frac{1}{s} \sum_{n=1}^\infty \widetilde F_n(s\,|\,L_0)\,.
\end{align}
We now substitute the expression \eqref{eq:fns} for the Laplace transform of
the $n^\text{th}$-passage probability and find that the Laplace transform of the
average number of encounters is
\begin{align}
\label{eq:avgn4}
  \langle \widetilde N(s) \rangle = \frac{1}{s}\sum_{n=1}^\infty
  \exp\left[-\epsilon \sum_{m=1}^{n-1}g_m(s)-L_0 \,g_n(s)\right]\,,
\end{align}
with $g_n(s)$ given in \eqref{eq:gns}.

\begin{figure}[h]
  \begin{center}
    \includegraphics[width=0.45\textwidth]{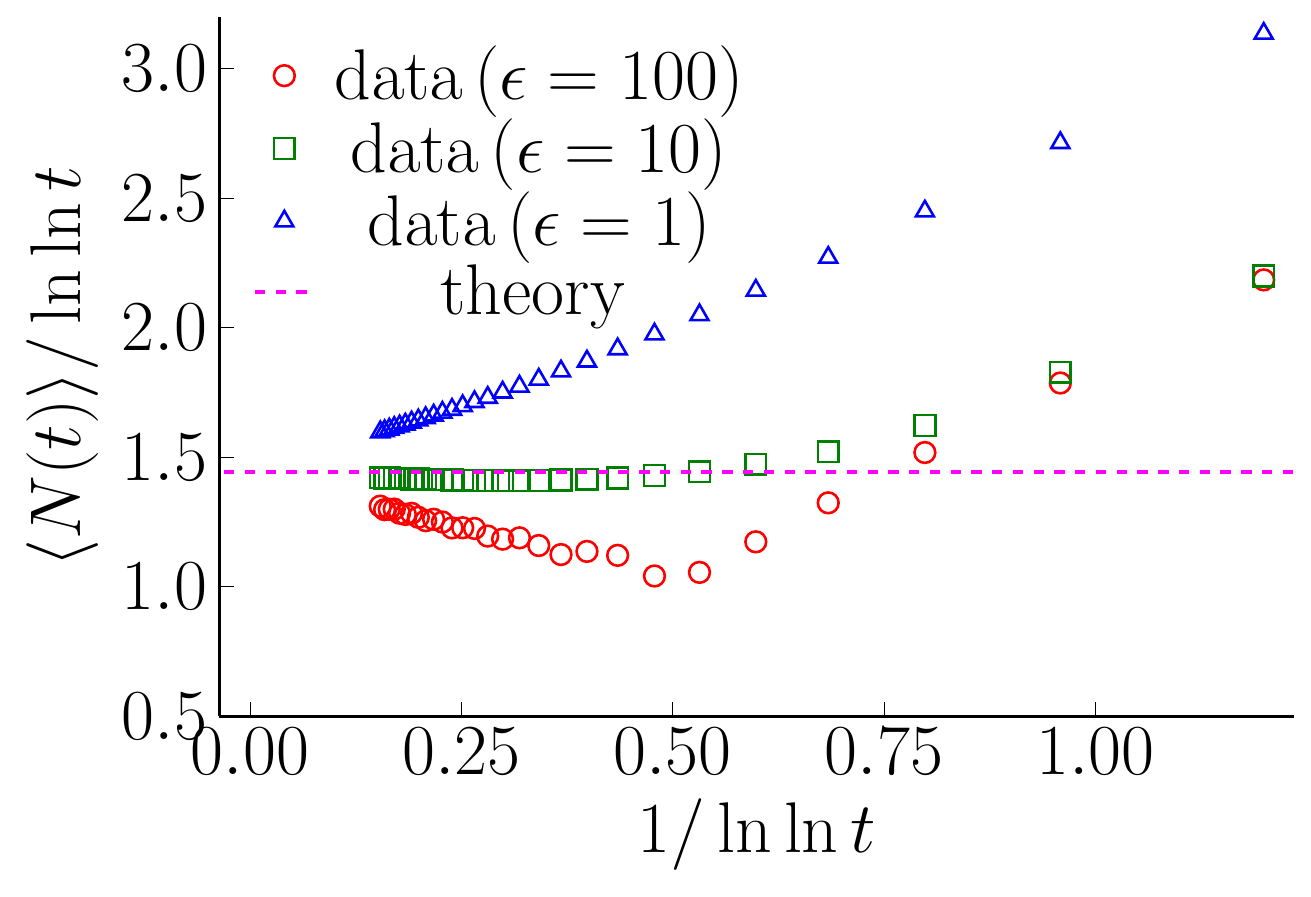}
    \caption{Average number of boundary encounters $\langle N(t) \rangle$
      scaled by $\ln\ln t$ as a function of the inverse of the scaled time for
      $D=L=\alpha=1$.  The theoretical prediction from \eqref{eq:Navgta} is
        also shown. }
    \label{fig:navg}
  \end{center}
\end{figure}

To obtain the long-time asymptotics, we focus on the $s\to 0$ limit of the
above expression:
\begin{align}
\label{eq:Navgsa1}
\langle \widetilde N(s) \rangle \sim \frac{1}{s}\sum_{n=1}^\infty
\exp\left[-\epsilon \sum_{m=1}^{n-1} \frac{\alpha^{1-2^{1-m}}}{D^{1-2^{-m}}}\,
  s^{2^{-m}}-L_0  \frac{\alpha^{1-2^{1-n}}}{D^{1-2^{-n}}}\,s^{2^{-n}}\right]\,. 
\end{align}
\end{subequations}
As shown in \ref{app:as}, this limiting behavior of
$\langle \widetilde N(s) \rangle$ reduces to
\begin{subequations}
\begin{align}
  \label{eq:Navgsa2}
  \langle \widetilde N(s) \rangle \sim \frac{1}{s \ln 2}\;\ln\left[\ln(\frac{1}{s})\right]\,,
\end{align}
which, in the time domain, gives the long-time behavior
\begin{align}
  \label{eq:Navgta}
\langle N(t)\rangle \sim \frac{1}{\ln 2}\ln\ln t\approx 1.4427\,\ln\ln t\,.
\end{align}
\end{subequations}
This dependence agrees well with numerical simulations for the number of
encounters shown in figure~\ref{fig:navg}.

\section{Location of the Boundary}

We now study the time dependence of the location of the boundary.  Let
$P(L,t)$ denote the spatial location of the boundary at time $t$.  This
probability distribution is given by
\begin{align}
    P(L,t) &= \delta(L-L_0)\;\text{erf}\left(\frac{L_0}{\sqrt{4\pi D t}}\right)\nonumber\\
& + \sum_{n=1}^{\infty}\int_0^{\infty} \!\!d\tau_1 \ldots d\tau_n \,
F_1(\tau_1\,|\,L_0) \times \ldots\times F(\tau_n\,|\,\alpha\,  \tau_{n-1}\!+\!\epsilon)\,
  \delta(L\!-\!L_0\!-\!n\epsilon \!-\!\sum_{m=1}^n\alpha \tau_m)\nonumber\\
 &\quad \times \text{erf}\left(\frac{\alpha\tau_n+\epsilon}{\sqrt{4\pi
   D(t-\sum_{m=1}^n \tau_m)}}\right)\Theta(t-\alpha\sum_{m=1}^n \tau_m)\,.
\end{align}
The first term accounts for the case where the diffusing particle never
reaches the boundary.  The $n^\text{th}$ term in the sum accounts for the
case where the particle reaches the boundary $n$ times.  Thus there must be a
first passage, a second passage,\ldots, up to an $n^\text{th}$ passage, after
which the particle cannot reach the boundary again.  These events are
accounted for by the product of $n$ first-passage probabilities and the
trailing error function.  The theta function imposes the condition that the
total time must be larger than the sum of the previous $n$ hitting times.  To
simplify notation in the formulas below, we introduce
$T_n\equiv \sum_{m=1}^n\tau_m$, the sum of the first $n$ time intervals
between successive encounters.

Computing the average value gives
\begin{align}
\langle L(t)\rangle &= L_0\,\text{erf}\left(\frac{L_0}{\sqrt{4\pi D t}}\right)\nonumber\\
 & + \sum_{n=1}^{\infty}\int_0^{\infty} d\tau_1 \ldots d\tau_n \,
F(\tau_1\,|\,L_0) \times \ldots \times F(\tau_n\,|\,\alpha\, \tau_{n-1}\!+\!\epsilon)\,
(L_0\!+\!n\epsilon \!+\!\alpha T_n)\nonumber\\
 &\quad \times \text{erf}\left(\frac{\alpha\tau_n+\epsilon}{\sqrt{4\pi
    D(t-T_n)}}\right)\Theta(t- T_n)\,.
\end{align}
It is now useful to Laplace transform the above relation.  This gives
\begin{subequations}
\begin{align}
\label{eq:Ls}
\langle \widetilde L(s)\rangle &= \frac{L_0}{s}\,
  \left(1-e^{-L_0 \sqrt{\frac{s}{D}}}\right)\nonumber\\[2mm]
 & +\frac{1}{s} \sum_{n=1}^{\infty}\int_0^{\infty} d\tau_1 \ldots d\tau_n \,
    F(\tau_1\,|\,L_0) \times\ldots\times F(\tau_n\,|\,\alpha\, \tau_{n-1}+\epsilon)\,
  (L_0+n\epsilon +\alpha T_n)\nonumber\\
 &\quad \times\left(1-e^{-(\alpha\tau_n+\epsilon)\sqrt{{s}/{D}}}\right)  e^{-sT_n}\,.
\end{align}
We argue that the main contribution to $ \langle \widetilde L(s)\rangle$
comes from the large-$n$ terms in the sum.  These terms will involve values
of $\tau_n$ that typically are also large.  Thus it is plausible that the
exponential term in the parentheses in the last line is negligible.  With
this assumption, we find
\begin{align}
  \langle \widetilde L(s)\rangle &\sim \frac{1}{s}
 \sum_{n=1}^{\infty}\int_0^{\infty}\!\! d\tau_1 \ldots d\tau_n \,
  F(\tau_1\,|\,L_0) \times\ldots\times F(\tau_n\,|\,\alpha\, \tau_{n-1}+\epsilon)\,
 (L_0\!+\!n\epsilon \!+\!\alpha T_n) e^{-sT_n}\,,
\end{align}
\end{subequations}
where we also drop the first term because it is negligible. Furthermore, the
integral of the 
$n$-fold product of first-passage probabilities is simply  $\widetilde
F_n(s\,|\,L_0)$.  Using this identification, we can write the average in the
simpler form
\begin{align}
  \langle \widetilde L(s)\rangle &\sim  
  \frac{1}{s}\left[\sum_{n=1}^{\infty}\left(L_0+n\epsilon- \alpha\,
   \frac{\partial}{\partial s}\right)\widetilde F_n(s\,|\,L_0) \right]\,.
\end{align}
Following similar reasoning as that given in \ref{app:as}, the leading-order
behavior is given by the last term in the brackets.  Thus
\begin{subequations}
\begin{align}
\langle \widetilde L(s)\rangle \sim
   -\frac{\alpha}{s}\frac{\partial}{\partial s}  \sum_{n=1}^{\infty}\widetilde F_n(s\,|\,L_0)
&\sim -\frac{\alpha}{s\ln 2}\frac{\partial}{\partial s}  \ln\left[\ln\left(\frac{1}{s}\right)\right]\,,\nonumber\\[1mm]
&\sim \frac{\alpha}{s^2 \ln 2 \ln(1/s)}\,,\qquad s\to 0\,.
\end{align}

 \begin{figure}[h]
  \begin{center}
    \includegraphics[width=0.5\textwidth]{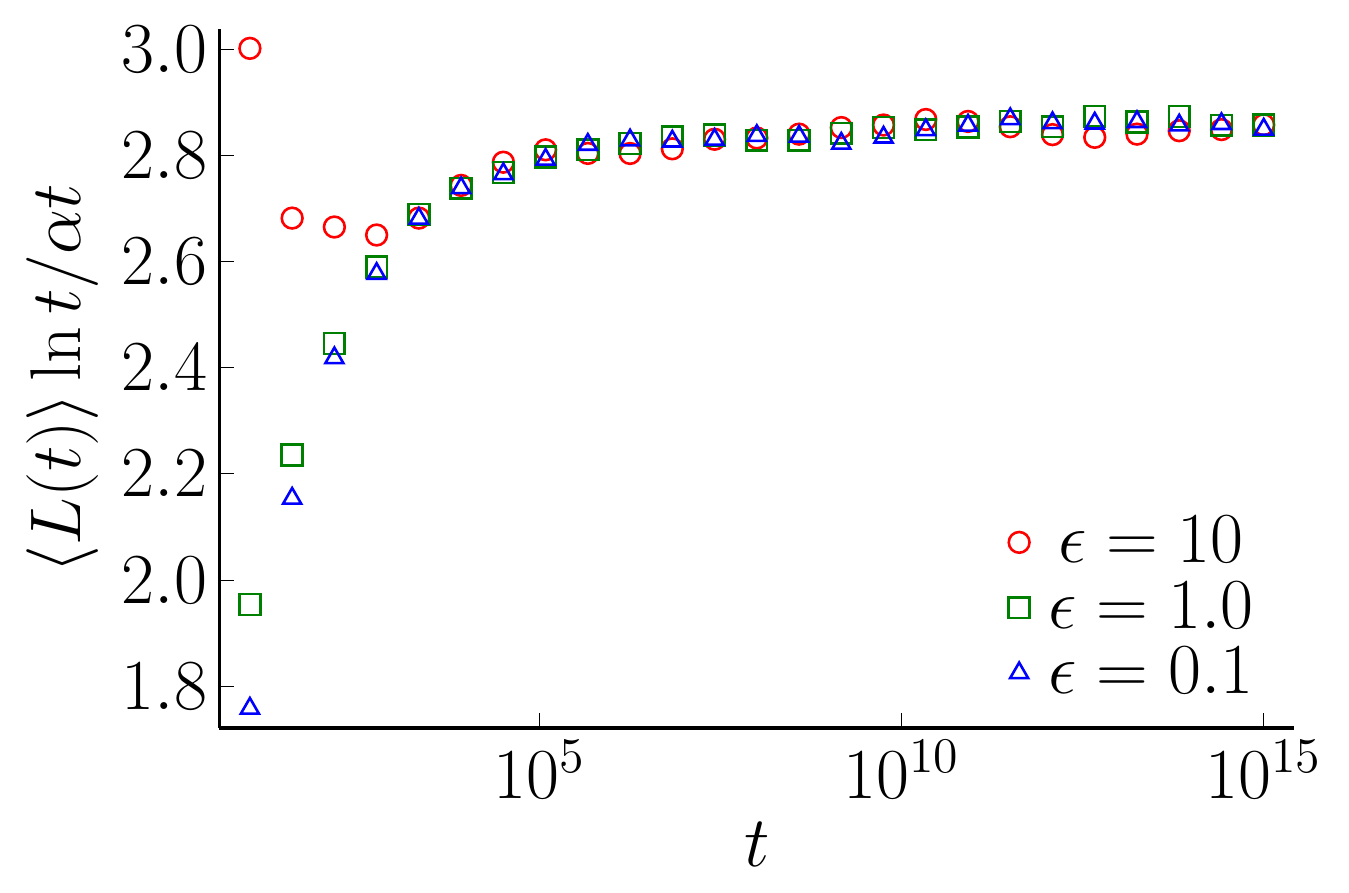}
    \caption{Rescaled average position of the boundary $\langle L(t) \rangle$
      as a function of time for $D=L=\alpha=1$.  }
    \label{fig:lavg}
  \end{center}
\end{figure}

In the time domain, we then obtain
\begin{align}
  \label{LT}
\langle L(t)\rangle \sim \frac{\alpha\, t}{\ln 2\ln t}\,,\qquad t\to \infty\,. 
\end{align}
\end{subequations}
This asymptotic behavior is in good agreement with numerical simulations as
shown in figure~\ref{fig:lavg}.  The numerical results clearly illustrate
that $\langle L(t) \rangle \propto t/\ln(t)$ for $t\to
\infty$. However, the amplitude that is found numerically is roughly twice
the value of the amplitude that is given in Eq.~\eqref{LT}.  We do not know
the source of the discrepancy, but it could stem from the neglect of the
exponential term in Eq.~\eqref{eq:Ls}.

\section{Concluding Comments}

We investigated a simple one-dimensional moving boundary problem that is
driven by the motion of a single Brownian particle.  This particle moves
freely on the infinite line and whenever it encounters the boundary, the
boundary instantaneously moves a distance that is proportional to the time
between successive collisions between the particle and the boundary.  We
determined some natural observables of this process.  The probability that
the particle \emph{first} hits the boundary for the $n^{\text{th}}$ time
asymptotically decays as $t^{-(1+2^{-n})}$.  Thus each successive
first-passage event is governed by a progressively fatter tail. The number of
collisions between the particle and the boundary scales as $\ln\ln t$; this
is the same dependence of the iterated logarithm law of free Brownian
motion~\cite{khintchine1924,feller2008introduction,morters2010brownian}, and
perhaps there is some unifying mechanism that links our moving boundary
problem with free diffusion.  In spite of the fact that encounters between
the particle and the boundary are rare, the position $L(t)$ of the boundary
moves nearly ballistically: $L(t)\sim t/\ln t$.  This rapid boundary motion
indicates that there must be some long time intervals between successive
particle-boundary encounters, so that the boundary moves by a large distance
when such an encounter occurs.

\section*{Acknowledgments}

BD acknowledges the financial support of the Luxembourg National Research
Fund (FNR) (App.~ID 14548297).  SR gratefully acknowledges partial financial
support from NSF Grant DMR-1910736.

\appendix
\section{Asymptotic analysis of $ \langle \widetilde N(s) \rangle$}
\label{app:as}

In this appendix, we study the small $s$ asymptotic behavior of
(\ref{eq:Navgsa1}):
\begin{align}
  \langle \widetilde N(s) \rangle \sim \frac{1}{s}
  \sum_{n=1}^\infty \exp\left[-\epsilon \sum_{m=1}^{n-1}
  \frac{\alpha^{1-2^{1-m}}}{D^{1-2^{-m}}}\;s^{2^{-m}}-L_0
  \frac{\alpha^{1-2^{1-n}}}{D^{1-2^{-n}}}\;s^{2^{-n}}\right]\,, \quad s\to 0\,. \label{eq:NavgsaA}
\end{align}
The sum over $n$ can be split roughly into two parts: (i) one is the
contribution when $n$ runs from $0$ to $n^*=n^*(s)$ such that
$s^{2^{-n^*}}=O(1)$, and (ii) the contribution when $n$ runs from $n^*$ to
$\infty$.  That is
\begin{align}
  \langle \widetilde N(s) \rangle \sim \frac{1}{s}\sum_{n=1}^{n^*}
  \exp\left[-L_0\frac{\alpha^{1-2^{1-n}}}{D^{1-2^{-n}}}\;s^{2^{-n}}\right]
  +  \frac{1}{s}\sum_{n^*}^\infty
  \exp\left[-\epsilon \left(\sum_{m=1}^{n-1} \frac{\alpha^{1-2^{1-m}}}{D^{1-2^{-m}}}\right)-L_0  \frac{\alpha^{1-2^{1-n}}}{D^{1-2^{-n}}}\right]\,, \quad s\to 0\,. \label{eq:NavgsaA2}
\end{align}
Because the series in the second term converges, we are left with 
\begin{align}
  \label{eq:NavgsaAs}
  \langle \widetilde N(s) \rangle \sim \frac{1}{s}\sum_{n=1}^{n^*}
  \exp\left[-L_0\frac{\alpha^{1-2^{1-n}}}{D^{1-2^{-n}}}\,s^{2^{-n}}\right]\,, \qquad s\to 0\,. 
\end{align}
Because the second term is negligible, the final result is independent of the
cutoff $\epsilon$.  The terms for $n<n^*$ will tend to $1$, and therefore we
find
\begin{align}
  \langle \widetilde N(s) \rangle \sim \frac{1}{s}\,n^*\,, \quad s\to 0\,.\label{eq:NavgsaAsz}
\end{align}
Finally, using that $s^{2^{-n^*}}=O(1)$, we obtain (\ref{eq:Navgsa2}).

\bibliographystyle{iopart-num}
\providecommand{\newblock}{}

\end{document}